\documentclass[12pt]{iopart}

\usepackage{graphicx}
\usepackage{epstopdf}

\begin{document}

\title{Bose-Einstein condensates of atoms with arbitrary spin}
\author{
P~Van~Isacker\dag\
and S~Heinze\ddag}

\address{\dag\
Grand Acc\'el\'erateur National d'Ions Lourds,
CEA/DSM--CNRS/IN2P3, BP~55027, F-14076 Caen Cedex 5, France}

\address{\ddag\
Institute of Nuclear Physics, University of Cologne,
Z\"ulpicherstrasse 77, 50937 Cologne, Germany}

\begin{abstract}
We show that the ground state of a Bose-Einstein condensate of atoms
with hyperfine spin $f=2$
can be either spin aligned,
condensed into pairs of atoms coupled to $F=0$,
or condensed into triplets of atoms coupled to $F=0$.
The complete phase diagram is constructed for $f=2$
and the generic properties of the phase diagram are obtained for $f>2$.
\end{abstract}






If atoms in a Bose-Einstein condensate (BEC)
are trapped by optical means~\cite{Stamper98},
their hyperfine spins (or spins) are not frozen in one particular direction
but are essentially free but for their mutual interactions.
As a result, the atoms do not behave as scalar particles
but each of the components of the spin
is involved in the formation of the BEC.
This raises interesting questions concerning the structure of the condensate
and how it depends on the spin exchange interactions between the atoms.

Such questions were addressed
in a series of theoretical papers by Ho and co-workers~\cite{Ho98}
who obtained solutions based on a generating function method.
In the case of spin-1 atoms the problem of quantum spin mixing 
was analyzed by Law {\it et al.}~\cite{Law98}
who proposed an elegant solution based on algebraic methods.
It is the purpose of this paper to point out
that a wide class of many-body hamiltonians
appropriate for the problem of interacting bosons with spin
can be solved through algebraic techniques
which have found fruitful applications in nuclear physics~\cite{Iachello87}
as well as in other fields of physics (see, {\it e.g.} Ref.~\cite{Iachello95}).
The main result derived in this paper
is that an exact solution is available for spin values $f=1$ and $f=2$
(for any value of the number of atoms $N$)
which allows the analytic determination
of the structure of the ground state of the condensate.
For spin values $f>2$
solvable classes of hamiltonians give insights
into the generic properties of the phase diagram.

We consider a one-component dilute gas of trapped bosonic atoms 
with arbitrary (integer) hyperfine spin $f$.
In second quantization the hamiltonian of this system
has a one-body and a two-body piece that can be written
as (we follow the notation of Ref.~\cite{Law98})
\begin{eqnarray}
{\cal H}&\equiv&{\cal H}_1+{\cal H}_2=
\sum_m\int
\hat\Psi_m^\dag
\left(-\frac{\nabla^2}{2M_{\rm a}}+V_{\rm trap}\right)\hat\Psi_m d^3x
\nonumber\\&&+
\sum_{m_i}\Omega_{m_1m_2m_3m_4}
\int\hat\Psi_{m_1}^\dag\hat\Psi_{m_2}^\dag\hat\Psi_{m_3}\hat\Psi_{m_4}d^3x,
\label{ham}
\end{eqnarray}
where $\hbar=1$, $M_{\rm a}$ is the mass of the atom,
and $\hat\Psi_m$ and $\hat\Psi_m^\dag$
are the atomic field annihilation and creation operators
associated with atoms in the hyperfine state $|fm\rangle$
with $m=-f,\dots,+f$,
the possible values of all summation indices in~(\ref{ham}).
The trapping potential $V_{\rm trap}$
is assumed to be the same for all $2f+1$ components.
The coefficients $\Omega_{m_1m_2m_3m_4}$
follow from the interaction between atoms
which is assumed to be of short-range, two-body character,
\begin{equation}
U(\vec x_i,\vec x_j)=
\delta(\vec x_i-\vec x_j)
\sum_{FM}\nu'_F|f^2;FM\rangle\langle f^2;FM|,
\label{inter}
\end{equation}
where $|f^2;FM\rangle$ is the combined state of the atoms $i$ and $j$
with total spin $F$,
and $\nu'_F\equiv4\pi\hbar^2a_F/M_{\rm a}$
with $a_F$ being the $s$-wave scattering length in the $F$ channel.
The assumption underpinning the form~(\ref{inter})
is rotational invariance of the hamiltonian in hyperfine-spin space.

We assume in this paper
that the scattering lengths in the different $F$ channels are comparable
and that, in first approximation, the interaction strength between the bosons
is independent of $F$.
In that case the dominant part of the hamiltonian~(\ref{ham})
is of the form
\begin{equation}
{\cal H}_{\rm s}=
{\cal H}_1+
\lambda'\sum_{m_1m_2}\int
\hat\Psi_{m_1}^\dag\hat\Psi_{m_2}^\dag\hat\Psi_{m_1}\hat\Psi_{m_2}d^3x,
\label{hamsym}
\end{equation}
and is symmetric under any interchange of the spin-component indices.
Under this assumption the condensate wave functions
for each spin component $\phi_m(\vec x)$ ($m=-f,\dots,+f$)
can be approximated by a single wave function $\phi(\vec x)$
which satisfies the Gross-Pitaevskii equation
associated with the dominant hamiltonian~\cite{Law98}.
Furthermore, the atomic field creation and annihilation operators at zero temperature
can be approximated by
\begin{equation}
\hat\Psi_m^\dag\approx b_m^\dag\phi(\vec x),
\quad
\hat\Psi_m\approx b_m\phi(\vec x),
\quad
m=-f,\dots,+f,
\label{fieldapp}
\end{equation}
where $b_m$ and $b_m^\dag$
are annihilation and creation operators associated with the entire condensate,
satisfying the usual boson commutation rules
\begin{equation}
[b_m,b_{m'}^\dag]=\delta_{mm'},
\qquad
[b_m,b_{m'}]=
[b_m^\dag,b_{m'}^\dag]=0.
\label{boscom}
\end{equation}
In this approximation the entire hamiltonian~(\ref{ham})
can be rewritten as
\begin{equation}
{\cal H}\approx\hat H\equiv
\epsilon\,b^\dag\cdot\tilde b+
{\frac 1 2}\sum_F\nu_F[b^\dag\times b^\dag]^{(F)}\cdot
[\tilde b\times\tilde b]^{(F)},
\label{hamgen}
\end{equation}
where the coefficients $\epsilon$ and $\nu_F$
are related to those in the original hamiltonian
through integration over $x$,
{\it viz.} $\nu_F=\nu'_F\int|\phi(\vec x)|^4d^3 x$.
The notation $\times$ in Eq.~(\ref{hamgen})
implies the coupling to a given spin $F$ and projection $M$,
\begin{equation}
[b^\dag\times b^\dag]^{(F)}_M=
\sum_{mm'}\langle fm\,fm'|FM\rangle
b^\dag_m b^\dag_{m'},
\label{angmom}
\end{equation}
where $\langle\cdot\cdot\,\cdot\cdot|\cdot\cdot\rangle$
is a Clebsch-Gordan coefficient~\cite{Edmonds57}.
Furthermore, the dot $\cdot$ denotes a scalar product,
\begin{equation}
\hat T^F\cdot\hat T^F\equiv
(-)^F\sqrt{2F+1}[\hat T^F\times\hat T^F]^{(0)}_0,
\label{scaprod}
\end{equation}
for tensor operators $\hat T^F_M$ of rank $F$.
The definition of the adjoint operator $\tilde b_m\equiv(-)^{f-m}b_{-m}$
ensures that $\tilde b_m$ is an annihilation operator
with transformation properties under rotations that are the same
as those for the creation operator $b_m^\dag$~\cite{Iachello06}.
With the above definitions we have
that $b^\dag\cdot\tilde b=\sum_m b_m^\dag b_m$
is the number operator $\hat N$
which counts the total number of atoms in the condensate.

To derive the solvability properties of the hamiltonian~(\ref{hamgen}),
we first determine its algebraic structure
by introducing the bilinear operators $b_m^\dag b_{m'}$.
From Eq.~(\ref{boscom}) one finds the commutation relations
\begin{equation}
[b_{m_1}b_{m_2}^\dag,b_{m_3}b_{m_4}^\dag]=
b_{m_1}b_{m_4}^\dag\delta_{m_2m_3}-b_{m_3}b_{m_2}^\dag\delta_{m_1m_4},
\label{comun}
\end{equation}
which can be identified
as those of the unitary (Lie) algebra U($2f+1$)~\cite{Iachello06}.
Exactly solvable hamiltonians with rotational or SO(3) invariance
are now found by the determination of all Lie algebras $G$
satisfying ${\rm U}(2f+1)\supset G\supset{\rm SO}(3)$.
The canonical reduction of U($2f+1$) is of the form
\begin{equation}
{\rm U}(2f+1)\supset{\rm SO}(2f+1)\supset{\rm SO}(3).
\label{chain}
\end{equation}
[For $f=3$ there is an additional exceptional ${\rm G}_2$ algebra
between SO($2f+1$) and SO(3)
which for the symmetric representations of U($2f+1$)
considered here does not add anything to the discussion.]
The relevance of a chain of nested algebras of the type~(\ref{chain})
is that it defines a set of commuting operators
and with it a class of solvable hamiltonians.
Consider in particular the hamiltonian
\begin{eqnarray}
\hat H'&=&
a_1\hat C_1[{\rm U}(2f+1)]+
a_2\hat C_2[{\rm U}(2f+1)]
\nonumber\\&+&
b\,\hat C_2[{\rm SO}(2f+1)]+
c\,\hat C_2[{\rm SO}(3)],
\label{hamsol}
\end{eqnarray}
where $a_1$, $a_2$, $b$, and $c$ are numerical coefficients
and $\hat C_n[G]$ is the $n^{\rm th}$-order Casimir operator of the algebra $G$
which satisfies the property
that it commutes with all generators of $G$~\cite{Wybourne74}.
Solvability of the hamiltonian~(\ref{hamsol}) follows from the fact
that it is written as a sum of commuting operators,
a property which indeed is valid for the Casimir operators
associated to any chain of {\em nested} algebras such as~(\ref{chain}).
The Casimir operators appearing in Eq.~(\ref{hamsol})
are known in closed form,
\begin{eqnarray}
\hat C_1[{\rm U}(2f+1)]&=&\hat N,
\nonumber\\
\hat C_2[{\rm U}(2f+1)]&=&\hat N(\hat N+2f),
\nonumber\\
\hat C_2[{\rm SO}(2f+1)]&=&
-(2f+1)\hat T^0_+\cdot\hat T^0_-+\hat N(\hat N+2f-1),
\nonumber\\
\hat C_2[{\rm SO}(3)]&=&
\sum_F
\left[{\textstyle{\frac 1 2}}F(F+1)-f(f+1)\right]\hat T^F_+\cdot\hat T^F_-
+f(f+1)\hat N,
\label{casimir}
\end{eqnarray}
in terms of the operators
$\hat T^F_{+,M}\equiv[b^\dag\times b^\dag]^{(F)}_M$
and $\hat T^F_{-,M}\equiv[\tilde b\times\tilde b]^{(F)}_M$.
Equations~(\ref{casimir}) show that the solvable hamiltonian~(\ref{hamsol})
is a special case of the general hamiltonian~(\ref{hamgen})
with coefficients $\epsilon$ and $\nu_F$
that are linear combinations of $a_1$, $a_2$, $b$, and $c$
according to
\begin{eqnarray}
\epsilon&=&a_1+(2f+1)a_2+2fb+f(f+1)c,
\nonumber\\
\nu_F&=&2a_2+2b+[F(F+1)-2f(f+1)]c,
\quad F\neq0,
\nonumber\\
\nu_0&=&
2a_2-4fb-2f(f+1)c,
\label{conditions}
\end{eqnarray}
The eigenvalues of the hamiltonian~(\ref{hamsol}) are
\begin{equation}
E'(N,v,F)=
a_1N+
a_2N(N+2f)+
b\,v(v+2f-1)+
c\,F(F+1).
\label{eigsol}
\end{equation}
The allowed values of $v$ are $v=N,N-2,\dots,1$ or 0,
as can be obtained from the ${\rm U}(2f+1)\supset{\rm SO}(2f+1)$
branching rule~\cite{Wybourne74}.
The quantum number $v$ corresponds to the number of bosons
{\em not} in pairs of bosons coupled to $F=0$,
and is known as seniority~\cite{Racah43,Talmi93}.
The allowed values of the total spin $F$
are obtained from the ${\rm SO}(2f+1)\supset{\rm SO}(3)$ branching rule
which is rather complicated but known in general~\cite{Gheorghe04}.
The $f=2$ example is discussed below.

The generic solvability properties of the original hamiltonian~(\ref{hamgen})
now follow from a simple counting argument.
For atoms with spin $f=1$
the solvable hamiltonian~(\ref{hamsol})
has three coefficients $a_1$, $a_2$, and $c$
[since SO($2f+1$)=SO(3)]
while the general hamiltonian~(\ref{hamgen})
also contains three coefficients $\epsilon$, $\nu_0$, and $\nu_2$.
[Note that the coupling of two spins to odd $F$
is not allowed in the approximation~(\ref{fieldapp})
of a common spatial wave function, so no $\nu_1$ term occurs.]
For atoms with spin $f=2$
both the solvable and general hamiltonian contains four coefficients
($a_1$, $a_2$, $b$, and $c$
{\it versus} $\epsilon$, $\nu_0$, $\nu_2$, and $\nu_4$)
which can be put into one-to-one correspondence.
Hence the general hamiltonian~(\ref{hamgen}) is solvable for $f=2$.
The same counting argument shows
that it is no longer solvable for $f>2$.

The case of interacting $f=1$ atoms was discussed by Law {\it et al.}~\cite{Law98}
who identified the existence of two possible condensate ground states:
one with all atoms aligned to maximum spin $F=N$
and a second with pairs of atoms coupled to $F=0$.
Whether the condensate is aligned or paired
depends on a single interaction parameter
which in our notation is $c$.
With the technique explained above we can also derive the phase diagram
for atoms with spin $f=2$.
The results are exact and valid for arbitrary $N$.
The entire spectrum is determined by the eigenvalue expression~(\ref{eigsol})
together with the necessary branching rules.
In particular, the allowed values of total spin $F$ for a given seniority $v$
are derived from the ${\rm SO}(5)\supset{\rm SO}(3)$ branching rule~\cite{Iachello87}
given by $F=2\tau,2\tau-2,2\tau-3,\dots,\tau+1,\tau$
with $\tau=v,v-3,v-6,\dots$ and $\tau\geq0$.

\begin{figure}
\includegraphics[width=7.5cm]{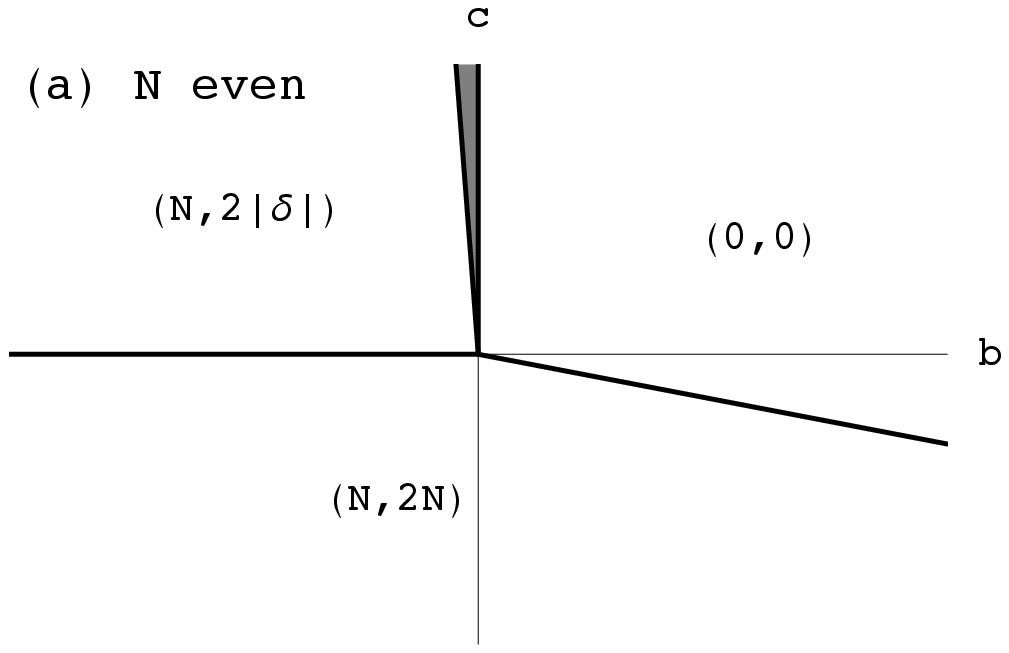}
\includegraphics[width=7.5cm]{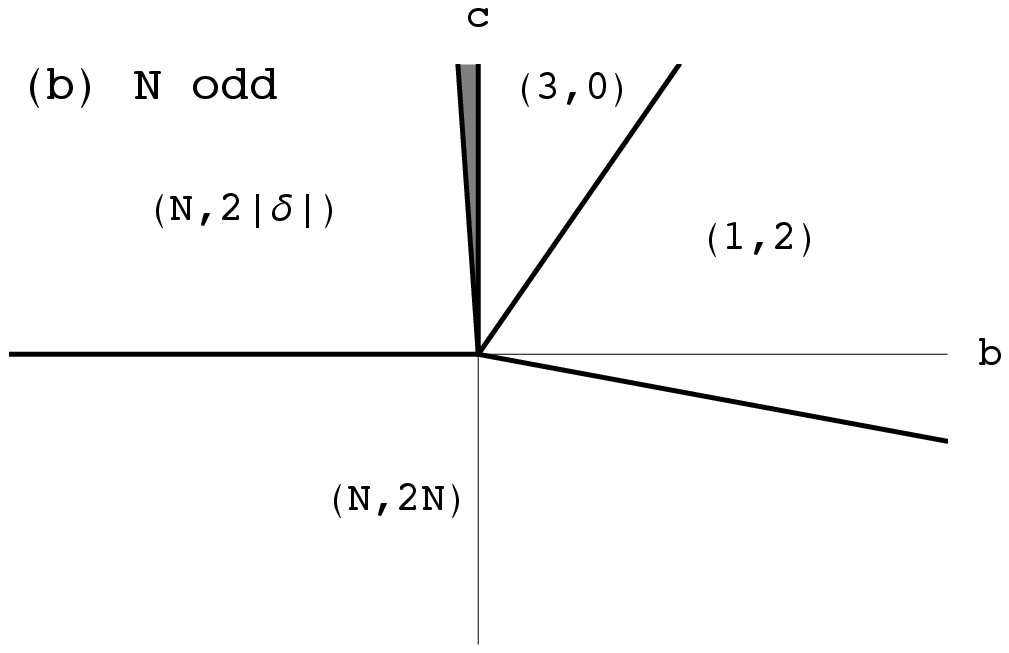}
\caption{\label{phase_eo}
Diagrams of the different phases
of a Bose-Einstein condensate of atoms with spin $f=2$
characterized by a ground state $(v_0,F_0)$
where $v_0$ is the seniority of the ground state and $F_0$ is its total spin.
The total number of atoms $N$ is even in (a) and odd in (b).
The grey area corresponds to a ground state with $(v_0,F_0)=(N-3+\delta,0)$
which only occurs for $\delta=\pm1$ and disappears in the limit $N\rightarrow\infty$.}
\end{figure}

It is now possible to determine
all possible ground-state configurations of the condensate.
This problem has been considered in the study
of the spectral features of quantal systems with random interactions~\cite{Chau02}.
We note that the character of the ground state
does not depend on the coefficients $a_i$
since the first two terms in the expression~(\ref{eigsol})
give a constant contribution to the energy of all states.
Although this contribution is dominant according to our earlier assumptions,
the spectrum generating perturbation of the hamiltonian
is confined to the last two terms
and depends solely on the coefficients $b$ and $c$
which are related to the original interactions $\nu_F$
according to
\begin{equation}
b=\frac{1}{70}(-7\nu_0+10\nu_2-3\nu_4),
\qquad
c=\frac{1}{14}(-\nu_2+\nu_4),
\end{equation}
The following exact finite-$N$ results are found
where the ground state of the condensate
is characterized by a seniority $v_0$ and a total spin $F_0$.
\begin{enumerate}
\item
$N$ is even.
We introduce $N=6k+2\delta$ with $k$ integer and $\delta=-1,0,+1$.
The possible ground-state configurations have
$(v_0,F_0)=(0,0)$, $(N,2N)$, $(N,2|\delta|)$, or $(N-3+\delta,0)$,
the latter existing only for $\delta=\pm1$,
\item
$N$ is odd.
We introduce $N=6k+3+2\delta$ with $k$ integer and $\delta=-1,0,+1$.
The possible ground-state configurations have
$(v_0,F_0)=(1,2)$, $(3,0)$, $(N,2N)$, $(N,2|\delta|)$, or $(N-3+\delta,0)$,
the latter existing only for $\delta=\pm1$.
\end{enumerate}
The phase diagram displays a richer structure than in the $f=1$ case
as is shown in Fig.~\ref{phase_eo}.
We observe first of all the presence of the aligned phase
where the seniority is maximal, $v_0=N$,
and all spins are aligned, $F_0=2N$.
Secondly, we have a low-seniority (paired) and consequently low-spin phase.
For even $N$ this corresponds necessarily to $(v_0,F_0)=(0,0)$.
For odd $N$ there must be at least one unpaired atom
leading to the ground-state configuration $(v_0,F_0)=(1,2)$;
alternatively, however, it might consist of a {\em triplet} of atoms
which is coupled to total spin $F_0=0$
leading to the ground-state configuration $(v_0,F_0)=(3,0)$.
The (1,2) and (3,0) phases are divided by the line $b=3c/7$.
The paired and aligned phases are separated by the line
\begin{equation}
b=-\frac{2N(2N+1)}{N(N+3)}c,
\qquad
b=-\frac{(2N-2)(2N+3)}{(N-1)(N+4)}c,
\end{equation}
for $N$ even or odd respectively,
which in both cases tends to $b=-4c$ for $N\rightarrow\infty$.

So far we have recovered the aligned and paired phases
also encountered for interacting $f=1$ atoms
(although the paired phase is somewhat more intricate for $f=2$
due to the possible presence of a triplet of atoms coupled to $F=0$).
For $f=2$ a third phase occurs for negative $b$ and positive $c$
characterized by high seniority ({\it i.e.} unpaired) and low total spin,
$(v_0,F_0)=(N,2|\delta|)$.
Finally, for $\delta=\pm1$ there exists a pathological region in the phase diagram
characterized by $(v_0,F_0)=(N-3+\delta,0)$ (see Fig.~\ref{phase_eo}).
It is separated from the high-seniority, low-spin region by the line 
\begin{equation}
b=-\frac{|\delta(\delta+3)|}{4(2N+\delta)}c,
\end{equation}
which tends to $b=0$ for $N\rightarrow\infty$.
Hence this region disappears in the large-$N$ limit.

We conclude that the ground state
of a BEC consisting of atoms with spin $f=2$
can be of three different types:
(i) a maximum-seniority spin-aligned,
(ii) a low-seniority low-spin,
or (iii) a maximum-seniority low-spin configuration.
Note that `seniority' in this context
refers to number of atoms that are not in {\em pairs} coupled to $F=0$.

Since the hamiltonian~(\ref{hamsol}) is solvable for $f=2$,
all eigenstates, and in particular the three different ground states,
can determined analytically.
The general expressions given by Chac\'on {\it et al.}~\cite{Chacon76} reduce to
\begin{eqnarray}
|v=N,F=M=2N\rangle&\propto&
\left(d_{+2}^\dag\right)^N|{\rm 0}\rangle,
\nonumber\\
|v=0,F=M=0\rangle&\propto&
\left(d^\dag\cdot d^\dag\right)^{N/2}|{\rm 0}\rangle,
\nonumber\\
|v=N,F=M=0\rangle&\propto&
\left([a^\dag\times a^\dag]^{(2)}\cdot a^\dag\right)^{N/3}|{\rm 0}\rangle,
\label{wave}
\end{eqnarray}
where the $f=2$ atoms are denoted as $d$ bosons.
In the second of these expressions it is assumed that $N$ is even
and in the third that $N=3k$;
other cases are obtained
by adding a single boson or an $F=0$ pair.
The $a^\dag$ are the so-called traceless boson operators~\cite{Chacon76}
which are defined as (see also Chapt.~8 of Ref.~\cite{Frank94})
\begin{equation}
a_m^\dag=d_m^\dag-\frac{d^\dag\cdot d^\dag}{2N+5}\tilde d_m.
\end{equation}
We emphasize that~(\ref{wave}) are the {\em exact} finite-$N$ expressions
for the eigenstates of the hamiltonian~(\ref{hamsol}).
Since in the large-$N$ limit the traceless boson operators $a_m^\dag$
become identical to $d_m^\dag$,
we arrive at a simple interpretation of the three types of configurations:
(i) spin-aligned,
(ii) condensed into {\em pairs of atoms} coupled to $F=0$,
and (iii) condensed into {\em triplets of atoms} coupled to $F=0$.

How will these features evolve with increasing spin $f$ of the atoms?
For arbitrary interaction strengths $\nu_F$ in the different $F$ channels
the hamiltonian~(\ref{hamgen}) is not solvable.
By imposing $f-2$ conditions on $\nu_F$
it can be brought into the form~(\ref{hamsol})
and this gives an idea of the structure of the general phase diagram
by constructing a two-dimensional slice of it.
For example, for atoms with spin $f=3$
the elimination of $a_1$, $a_2$, $b$, and $c$ from Eq.~(\ref{conditions})
yields the condition $11\nu_2-18\nu_4+7\nu_6=0$.
For $f>3$ more conditions on $\nu_F$ are found.
If all conditions are satisfied,
the phase diagram in $b$ and $c$ with
\begin{equation}
b=\frac{-7\nu_0+10\nu_2-3\nu_4}{14(2f+1)},
\qquad
c=\frac{1}{14}(-\nu_2+\nu_4),
\end{equation}
has properties similar to those in the $f=2$ case.
The analysis requires the knowledge of the multiplicity $d^{(f)}_v(F)$,
({\it i.e.}, the number of spin-$f$ atom states
with seniority $v$ coupled to total spin $F$)
which can be derived
from the ${\rm SO}(2f+1)\supset{\rm SO}(3)$ branching rule~\cite{Gheorghe04}.
We find that for sufficiently large even $N$ there are four competing ground states
with $(v_0,F_0)=(N,fN)$, $(N,0)$, $(0,0)$, and $(2,2)$,
the latter of which disappears as a ground state in the large-$N$ limit.
For sufficiently large odd $N$ the four competing ground states have
$(v_0,F_0)=(N,fN)$, $(N,0)$, $(1,f)$, and $(3,f_2\equiv f\bmod2)$,
the latter two being separated by the line
$b=[f(f+1)-f_2(f_2+1)]c/(4f+6)$.
The results correspond to what is found in the $f=2$ case
and lead to an essentially identical $(b,c)$ phase diagram.

Finally, we point out that the appearance of exact seniority ground states
requires weaker conditions on $\nu_F$
than those that have been discussed so far.
In fact, the spin-aligned configuration $(N,fN)$
is always an eigenstate of the general hamiltonian~(\ref{hamgen})
because the $F=fN$ state is unique.
Furthermore, it can be shown~\cite{Talmi93}
that seniority is a good quantum number 
if the interaction strengths $\nu_F$
satisfy $\lfloor f/3\rfloor$ conditions only
(where $\lfloor x\rfloor$ is the largest integer
smaller than or equal to $x$).
For all cases of any conceivable interest for BECs,
this reduces to no condition on the strengths $\nu_F$ for $f=1,2$
or just a single one for $f=3,4,5$.
So there is at most a single condition required
for all eigenstates to carry exact seniority
and for the results of this paper to be valid.
Nevertheless, the determination of the complete phase diagram for $f>2$
with unconstrained interaction strengths $\nu_F$
remains a problem worthy of further investigation.

\section*{Acknowledgments}
We acknowledge a conversation with H.T.~Stoof
that prompted our interest in this problem.

\end{document}